\documentclass[square]{ws-procs11x85}
\usepackage{balance} %necessary when used with \twocolumn
\usepackage{amssymb}

\def\Journal#1#2#3#4{{#1} {\bf #2}, #3 (#4)}

\def\be{\begin{equation}}
\def\ee{\end{equation}}
\def\bc{\begin{center}}
\def\ec{\end{center}}
\def\beq{\begin{eqnarray}}
\def\eeq{\end{eqnarray}}
\def\d{{\rm d}}
\def\doppler{\delta}
\def\betaj{\beta_{\rm j}}
\def\Gammaj{\Gamma_{\rm j}}

\def\gmax{\gamma_{\rm \max}}
\def\gmin{\gamma_{\rm \min}}
\def\Gammaf{\Gamma_{\rm f}}

\def\msun{{M_\odot}}

\def\taugg{\tau_{\gamma\gamma}}

\def \Tmax{T_{\max}}
\def \Thetamax{\Theta_{\max}}
\def\etaiso{\eta_{\rm d}} 

\def\Lj{L_{\rm j}}
\def\LB{L_{\rm B}}
\def\Ld{L_{\rm d}}
\def\Ldfv{L_{\rm d,45}}
\def\Rj{R_{\rm j}}

\def\ergs{\rm erg\ s^{-1}}

\def\me{m_{\rm e}}

\newcommand{\met}{\hbox{E\kern-0.5em\lower-0.1ex\hbox{/}}_T}

\begin{document}

\twocolumn[
\title{Photon breeding mechanism in relativistic jets: astrophysical implications}

\author{J. Poutanen$^{1}$ and B. E. Stern $^{2,3,1}$}

\address{$^{1}$ Astronomy Division, Department of Physical Sciences, P.O.Box 3000, 
90014 University of Oulu, Finland \\ E-mail:  juri.poutanen@oulu.fi}

\address{$^{2}$ Institute for Nuclear Research, Russian Academy of Sciences,
Prospekt 60-letiya Oktyabrya 7a, Moscow 117312, Russia} 

\address{$^{3}$ Astro Space Center, Lebedev Physical Institute, Profsoyuznaya 84/32,  Moscow 117997, Russia \\
E-mail: stern@bes.asc.rssi.ru}

%%%%%%%%%%%%%%%%%%%%%%%%%%%%%%%%%%%%%%%%%%%%%%%%%%%%%%%%%%%%%%%%%%%%%%%%%
% You may repeat \author \address as often as necessary                 %
%%%%%%%%%%%%%%%%%%%%%%%%%%%%%%%%%%%%%%%%%%%%%%%%%%%%%%%%%%%%%%%%%%%%%%%%%

\begin{abstract}
Photon breeding in relativistic jets involves multiplication of high-energy photons propagating 
from the jet to the external environment and back with the conversion into 
electron-positron pairs. The exponential growth of the energy density of these photons is a 
super-critical process powered by the bulk energy of the jet. 
The efficient deceleration of the jet outer layers creates a structured jet morphology 
with the fast spine and slow sheath. In initially fast and  high-power jets even the spine can be 
decelerated efficiently leading to very high radiative efficiencies of conversion of the jet bulk
energy into radiation. The decelerating, structured jets have angular distribution of radiation 
significantly  broader than that predicted by a simple  blob model with a constant Lorentz factor. 
This reconciles the discrepancy between the high Doppler factors determined by the fits to
the spectra of TeV blazars and the low apparent velocities observed at VLBI scales 
as well as the low jet Lorentz factors required by  the observed statistics and luminosity ratio 
of Fanaroff-Riley I  radio galaxies and BL Lac  objects.
Photon breeding produces a population of high-energy leptons  in agreement with the 
constraints on the electron injection function required by spectral fits of the TeV blazars. 
Relativistic pairs created outside the jet and emitting gamma-rays by inverse Compton process might 
explain the relatively high level of the TeV emission from the misaligned jet in the radio galaxies.
The  mechanism reproduces basic spectral features observed in blazars including
the blazar sequence (shift of the spectral peaks towards lower energies with increasing luminosity).
The mechanism is very robust and can operate in various environments characterised by the 
high photon density.  
 \end{abstract}
\vskip12pt  % insert '\vskip12pt' while using '\twocolumn' command
%\vskip28pt % if there is no keywords
]

\bodymatter

\section{Introduction}

A general consensus exists on the radiative processes responsible for the blazar emission: synchrotron, synchrotron self-Compton (SSC) and external radiation Compton (ERC) mechanisms. However,  neither  the distance from the central black hole to  the gamma-ray emitting region, nor the mechanisms of the energy dissipation and electron acceleration in relativistic jets are well understood. The internal shocks model \cite{spa01}  as well as magnetic  reconnection models \cite{lut03} have been discussed as possible scenarios for  the energy dissipation mechanism. 

The arguments based on the gamma-ray transparency and the observed short time-scale variability  constrain the gamma-ray emitting region in blazars to distances of about $\sim10^{17}$ cm \cite{s94}. This distance is  often associated with the broad-emission line region (BLR). In internal shock model, there is no physically sound way of explaining this distance. In addition,  the internal shocks are rather  inefficient radiators, unless huge fluctuations of the Lorentz factors are involved \cite{amb00}. Very rapid TeV variability detected from PKS 2155--304 \cite{aha07}  and Mrk 501 \cite{alb07} requires small distances and/or small emission region size and makes the internal shock models questionable. The  reconnection models are not yet at the level to predict the distance or efficiency. 
 
Recently, a novel photon breeding  mechanism for dissipation of the bulk energy of relativistic jets was suggested \cite{SP06,SP08}. The mechanism  involves multiplication of high-energy photons propagating from the jet to the external environment and back with the conversion into electron-positron pairs. The exponential growth of the energy density in these photons is a super-critical process  \cite{SP08b} powered by the jet bulk energy. The general scheme of the converter mechanism, where conversion of charged particles to neutral ones and back plays an important role in particle acceleration, was first discussed independently in \cite{der03,st03}. The numerical studies of the  operation of the photon breeding mechanism in ultrarelativistic shocks \cite{st03}, showed  the high radiative efficiency of the bulk energy dissipation.   A similar problem for relativistic jets in active galactic nuclei was investigated by us 
\cite{SP06,SP08}  using a ballistic model of the jet and  a detailed treatment of particle propagation and interactions. We showed that a large fraction of the jet bulk energy can be converted into gamma-rays, if the following  conditions are met: jet Lorentz factor $\Gammaj \gtrsim 5$--10,  the presence of the ambient  soft photon field  and a transversal or a chaotic magnetic field in the jet and the ambient medium.

In this paper we discuss the observational predictions of the photon breeding mechanism and compare them to the available data on blazars and radio galaxies. In Section \ref{sec:sites}  we discuss the possible sites, where photon breeding can operate. In Section \ref{sec:jet} we describe the character of the jet radiation in the photon breeding regime, particularly, the angular distributions of the emission components. We also show that it naturally resolves the Doppler factor crisis.  In Section \ref{sec:spectr} we compare numerical simulations to the observed broad-band blazar spectra as well as to the electron  distributions obtained from the spectral fits.

\section{Photon breeding operation sites} 
\label{sec:sites} 

The operation of the photon breeding mechanism  requires relativistic motion with at least $\Gammaj>4$ \cite{SP06} and a source of sufficiently dense transverse photon field (e.g. BLR). The photon breeding thus allows various emission sites. The minimum requirement is that the optical depth for the high-energy photons to pair production on soft photons is not much smaller than unity across the jet. If the background of soft photons is given by the multi-colour accretion disc of luminosity $\Ld$, this condition can be written as \cite{SP06}\footnote{We use standard notations $Q=10^x Q_x$ in cgs units and for dimensionless variables.} 
\be \label{eq:taugg}
\taugg(R) = 60 \frac{\Ldfv}{R_{17} \Theta_{\max,-5}} (10\theta) \gtrsim 1 ,
\ee
where  $\Thetamax\equiv k\Tmax/ \me c^2 \approx 10^{-5}$ is the maximum disc temperature, $R$ is the distance to the emission site, $\theta=\Rj/R$ is opening angle of the jet. 

At distances comparable to the size of the accretion  disc, its direct radiation can serve as a target for high-energy photons and can  trigger the photon avalanche, if the jet is already accelerated there. The photon-photon opacity through the jet is very large at these distances and the jet deceleration and radiative efficiency might be rather small, because only a very thin boundary layer can 
participate in photon breeding.  The presence of the X-rays from the accretion disc  extending up to $\sim 100$ keV can be important and can trigger a pair cascade at small distances \cite{Lev96}.

At larger distances the disc radiation becomes more beamed along the jet and the disc photons 
do not interact anymore with the high-energy radiation produced in the jet. 
The BLR radiation, however, supplies enough soft photons to satisfy  the  conditions for photon breeding. 

At the parsec scale, if the jet is still relativistic enough,  the process can operate on the IR radiation from the dusty torus. 
Assuming a black body dust emission  with the temperature 
$\Theta_{\rm dust}=kT_{\rm dust}/\me c^2 \approx 10^{-7} \Ldfv^{1/4} R_{\rm pc}^{-1/2}$, 
the  optical depth is about  \cite{SP08}
\be \label{eq_tauggdust}
\taugg  =  60 \etaiso (10\theta) \Ldfv^{3/4} R_{\rm pc}^{-1/2}   ,
\ee
where  $\etaiso$ is the ratio of the dust to the disc luminosities.  
Thus, even for $\etaiso$ of the order of a few percent, the photon breeding
is effective at the parsec scale in bright quasars. 

Even further out, if jet is still highly relativistic, 
photon breeding might operate on the stellar radiation field  at kpc scale 
and  at $\sim$100 kpc on the cosmic microwave background radiation. 

Finally, we would like to note, that the emission from the blazar gamma-ray zone is very bright 
within the jet opening angle and can strongly affect 
the conditions for photon breeding at larger distances, where the radiation of the 
produced pairs can be responsible for the jet emission observed in radio galaxies.

%######################################
\begin{figure}
\centerline{\epsfig{file=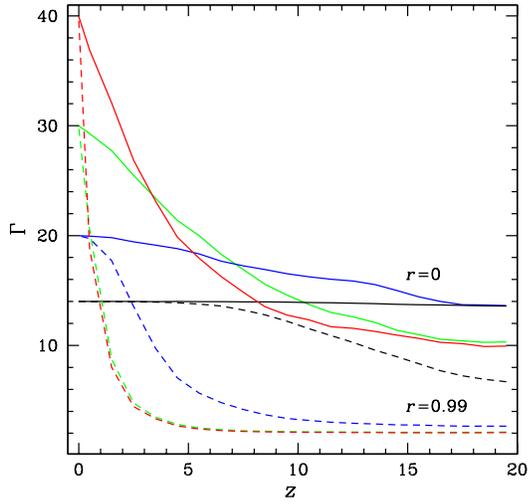,width= 7truecm}}
\caption{Distribution of the fluid Lorentz factor in the direction of jet propagation (measured in units 
of radius of the jet cross-section $\Rj= \theta R$)  for various starting jet Lorentz  factors: $\Gammaj=14$, 20, 30, 40.  
Solid curves show the Lorentz factor at the jet axis $r=0$, and the 
dashed curves are for the jet boundary at $0.99\Rj$. 
Parameters of the simulations: disc luminosity $\Ld=3\times 10^{44}\ergs$, 
jet power $\Lj=\Ld$, Poynting flux $\LB=0.2\Ld$, distance to the BLR
$R=10^{17}$ cm, jet opening angle $\theta=0.05$. }
\label{fig:gamz}
\end{figure}
 %######################################

\section{Angular distribution, Doppler factor crisis and volume dissipation} 
\label{sec:jet}

\subsection{Emission pattern in photon breeding model}

Photon breeding in the relativistic shear layers causes efficient momentum exchange between 
rapidly moving outer layers of the jet and the external environment \cite{SP06,SP08}. 
ERC scattering by the relativistic electron-positron pairs born within the jet, being rather anisotropic 
in the jet comoving frame, not only produces gamma-ray emission, but also carries away a fraction 
of the jet momentum. This causes first deceleration of the  jet's outer layers (see dashed curves in
Fig. \ref{fig:gamz}). As the cascade develops within the jet, even the spine can then 
be decelerated (see solid curves in Fig. \ref{fig:gamz})

The main radiative mechanisms responsible for the jet emission are the same as usually assumed to describe blazar spectra: synchrotron, SSC, and ERC. In a differentially decelerating jet, the synchrotron radiation from the slower sheath can dominate the energy density of soft photons within the faster spine. Thus  the energy loss may be dominated by the 
external synchrotron Compton (ESC) mechanism (see also \cite{gk03, gtc05}). 

\begin{figure}[t]
\center
\centerline{\psfig{figure=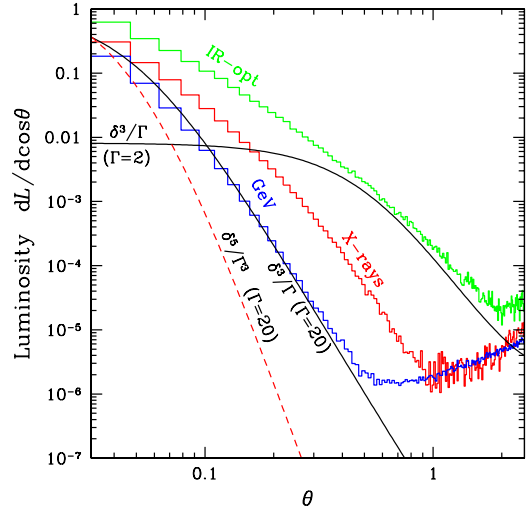,width=7truecm}}
\caption{Angular distribution of luminosity  in the three energy bands (IR-optical, X-rays and GeV gamma-rays)
from the jet differentially decelerating under action of photon breeding
(histograms, arbitrary normalization).  
Parameters:  $\Gammaj=20$, $\Ld= \Lj= \LB= 3\times 10^{45}\ergs$, $R=10^{17}$ cm, $\theta=0.05$.
Theoretical dependencies  expected for the steady relativistic jet emitting 
isotropically in the comoving frame (\ref{eq:isojet}) 
 and the  ERC radiation (\ref{eq:erciso}) are shown by solid curves (cases with $\Gamma=2$ and 20) and 
dashed curve, respectively.
}
\label{fig:angle}
\end{figure}

The angular distribution of radiation  from the differentially decelerating jet is very different from that 
predicted by simplistic models invoking relativistically moving blobs. 
For an isotropically emitting blob, the angular pattern  is given by 
\be \label{eq:isojet}
A(\theta) = \frac{\doppler^3}{\Gammaj} ,
\ee
where $\doppler=1/\Gammaj(1-\betaj\cos\theta)$ is the Doppler factor (see solid curves in Fig. \ref{fig:angle}), 
while the ERC beaming is given by  \cite{d95,sso03,SP08} (dotted curve in Fig. \ref{fig:angle})
\be \label{eq:erciso}
A(\theta) =    \frac{3}{4} \frac{\doppler^5}{\Gammaj^3} .
\ee
Photon breeding gives a much wider beams.  
The histograms in Fig.~\ref{fig:angle} show the angular dependence 
of the luminosity in three typical energy bands  (IR-optical, X-rays and gamma-rays).

We see that the distribution of high-energy photons at small angles
follows quite closely  equation (\ref{eq:isojet}).   
The low-energy (IR-optical  and the X-rays) photons are much less beamed.  
The spatial gradients of $\Gamma$ is the main cause of that. 
The soft band is dominated by the synchrotron radiation and 
the emission at $\theta=0.5$ exceeds the simple estimate  (\ref{eq:isojet}) 
by 3 orders of magnitude.

As a result of photon breeding, a significant population of relativistic pairs is also born in the external environment. 
They inverse Compton scatter the disc and BLR radiation  and synchrotron photons from the jet 
more or less isotropically producing an additional component clearly visible at large angles.

\subsection{Doppler factor crisis in TeV blazars}

The Doppler factors required by the homogeneous SSC models to describe the spectra of the 
blazars emitting at TeV energies (to avoid absorption by the infrared radiation), 
are large $\doppler$$\sim$20--100  \cite{gcc02, kca02,kon03,gdk07}.  
On the other hand,   the apparent velocities observed at the parsec scale 
in  TeV blazars \cite{mar99,pe04,pe05} are mildly relativistic. 
In addition, the jet Lorentz factors, derived by matching the luminosity functions 
and statistics of BL Lac objects and their parent population of FR I radio galaxies, 
was estimated to be $\Gammaj \sim$3--6 \cite{up95, ccc00,har03}. 
Such a clear disagreement between  the jet Lorentz factors determined by various methods 
has prompted to talk about the Doppler factor crisis \cite{tav05}.

One proposed solution to this crisis is that the jet Lorentz factor   
drops from 20--50 at the gamma-ray emitting, subparsec scale to just a few at 
the radio-emitting VLBI parsec scale  \cite{gk03}. 
For the viewing angle within the beaming pattern of the initial jet, $\theta\lesssim 1/\Gammaj$, 
the apparent velocity of the decelerated jet with Lorentz factor $\Gammaf\ll\Gammaj$  is \cite{SP08}
\be
\beta_{\rm f, app} \approx 2 \theta \Gammaf^2 \lesssim 2 \frac{\Gammaf^2}{\Gammaj} . 
\ee
Thus, deceleration of the jet to $\Gammaf< \sqrt{\Gammaj/2}$ guaranties 
that the apparent motion is subluminal. 

It was also suggested that the jet may consists of  a fast spine and a slow sheath (so called structured jet) \cite{ccc00,gtc05}. In this model a slower sheath dominates the emission in the off-axis sources. 

However, both proposals lack  a physical mechanism for dissipation of a significant fraction of the jet bulk energy to achieve a required deceleration. The photon breeding mechanism, on the other hand,  produces a decelerating and structured jet in a self-consistent way   (Fig. \ref{fig:gamz}; \cite{SP08,SP08b}),  and therefore it is capable of unifying BL Lacs with radio galaxies keeping high $\Gammaj$ to explain the gamma-ray emission and  may  resolve thus the Doppler factor crisis.

\subsection{Off-axis emission and the TeV emission from radio galaxies}

As we have discussed above, the photon breeding mechanism naturally produces very broad beams of photons (see also \cite{der03,der07}). Interestingly, the ratio of soft to hard photon luminosities in not a monotonic function of the viewing angle  (see Fig. \ref{fig:angle}). In slightly misaligned objects the gamma luminosity may be very weak compared to the optical luminosity, while again in radio galaxies observed at large angles $\theta\sim1$,  the relative contribution of the gamma-ray flux increases due to the  emission by the externally produced  pairs.   The luminosity ratio between the nearly isotropic emission and the beamed emission at $\theta\approx1/\Gammaj$    is about $\Gammaj^{-4}$   (two powers of $\Gammaj$ come from the energy amplification in the jet and two powers appear because of beaming into a solid angle $\sim 1/\Gammaj^2$). This ratio is $\sim\Gammaj^2/10$ larger than that predicted by equation (\ref{eq:isojet}). Thanks to the emission from the external medium, the nearby misaligned  jets become observable in high-energy gamma-rays. 

The central source of M87, which has the best studied jet, has 
a rather low disc luminosity of only  $\sim10^{42} \ergs$ \cite{bsh91}.
Sufficient pair-production opacity (\ref{eq:taugg}) can be achieved at $R < 10^{16}$ cm, 
which is only $\sim 10$ Schwarzschild radii for the estimated mass of the central black hole $\sim 3 \cdot 10^9 \msun$   
\cite{mac97}. At such a distance the photon breeding can operate on the direct disc radiation. 
The implied distance is consistent with the detection of the 
rapid  ($\sim10^5$ s) variability of the TeV photon flux from M87 \cite{aha06}.

\subsection{Observational appearance and volume dissipation}
 
Most of the emission from the jet undergoing photon breeding comes 
from the regions of maximum gradient of $\Gamma$. In the steady-state, 
the volume emissivity contributing to the observed emission (at small angles to the jet) is 
 \be
j(r,z) \propto \frac{\d \Gamma(r,z)}{\d z} \frac{\doppler^3}{\Gamma} . 
\ee
In the case of small luminosities (and small $\Gammaj$) only the outer layers of the jet
suffer significant deceleration, resulting in the limb-brightening,
with most of the emission coming from large $z$, where the cascade had time to develop fully. 
The situation is more complicated  at high luminosities (or large  $\Gammaj$). 
The outer layers decelerate rapidly (see curves for $\Gammaj=30$, 40 in Fig. \ref{fig:gamz}) 
making the emission limb-brightened at small $z$, while the layers close to the jet axis decelerate later. 
Thus the emission from the core dominates  at large $z$. 
At angles larger than $\gtrsim 1/\Gammaj$, only slower, significantly  decelerated layers contribute 
to the observed luminosity, and therefore we always would see the limb-brightened emission. 
 
Our predictions are consistent with the observed limb-brightened  morphology of the Mrk 501
jet at the parsec scale  \cite{gir04}. 
Such a  structure  could  a result from the efficient deceleration and  loading by relativistic pairs 
of the jet outer layers by the photon breeding  at subparsec scale. 
 
In spite of the fact, that deceleration is more efficient at the jet boundary, 
the energy dissipation can involve the whole volume of the jet as it is defined by 
the mean-free path of high-energy photons. 
This is in a striking contrast to the Fermi-type diffusive acceleration,
where only a narrow boundary layer  is active, because of the much smaller Larmor radii.

\section{Spectral energy distribution of blazars and the electron distribution}
 \label{sec:spectr}
 
As a result of photon breeding, a population of high-energy electron-positron pairs is produced. The pair injection spectrum should be bounded and mirror (relative to $\me c^2$) the spectrum of the soft photon background, because the pairs in the jet are produced only 
by those high-energy photons that can interact with the soft photons.  For example, if soft photons are from the accretion disc of maximum temperature $\Thetamax$, the low-energy injection cutoff in the jet frame is at $\gmin \sim \Gammaj/6\Thetamax$ \cite{SP06,SP08}. This cutoff should depend weakly  on the luminosity and the black hole mass, which defines the characteristic emission radius, so that $\Thetamax\approx 10^{-5}\Ldfv^{1/4}(M_{\rm BH}/10^8\msun)^{-1/2}$ and thus we can predict that 
\be \label{eq:gmin}
\gmin \sim 3\times 10^5 \left(\frac{\Gammaj}{20}\right) \Ldfv^{-1/4} 
\left( \frac{M_{\rm BH}}{10^8\msun}\right) ^{1/2}. 
\ee
At high luminosities, however, the synchrotron photons from the jet can provide enough opacity  and $\gmin$ may be smaller. 

%######################################
\begin{figure}
\centerline{\epsfig{file=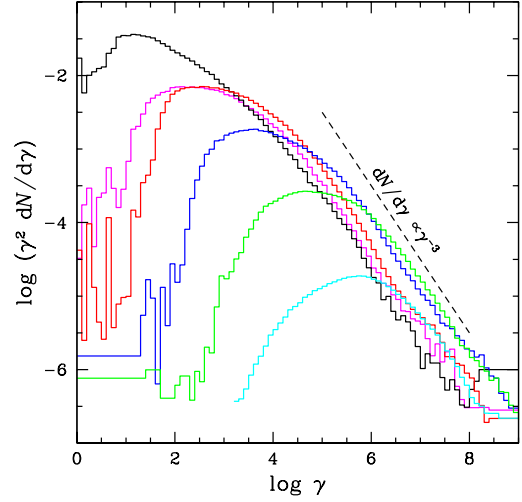,width=7truecm}}
\caption{Electron (and positron) energy distributions 
in the jet comoving frame (averaged over the jet volume) as a result of photon breeding. 
Simulation correspond to $\Gammaj=20$,   $\Lj = \Ld$,  $\LB=0.2\Ld$, with 
$\Ld$ varying  (from bottom to top histograms) from $5\times10^{43}$ to $10^{46}$ erg s$^{-1}$.
 }
\label{fig:espec}
\end{figure}
%######################################

The detailed fitting of the broad-band spectra of TeV blazars and low-power BL Lacs with the SSC model shows that the distribution of relativistic electrons injected to the system can be modelled as a power-law of index $p\approx 2$--2.5 with the low-energy cutoff at $\gmin \sim 10^4$--$10^5$ and the high-energy cutoff at $\gmax\sim 10^6$--$10^7$ \cite{gcc02, kca02,kon03,gdk07}.\footnote{The high-luminosity objects require  much lower electron energies \cite{gcc02}, but this result may be biased because the electron distribution is obtained from one-zone models, while the synchrotron peak in reality may not be related to the gamma-rays.} Such a peaked distribution of injected electrons is difficult, if not impossible, to obtain with the standard Fermi-type acceleration mechanisms, while the photon breeding can reproduce it easily (see Fig. \ref{fig:espec}).

At higher luminosities (and compactnesses), the high-energy photons produce a pair cascade. 
This reduces the mean energy per pair (see Fig. \ref{fig:espec}). At these compactnesses the pairs also cool more efficiently producing a cooling distribution $\d N/\d \gamma\propto \gamma^{-2}$, which changes to a steeper $\gamma^{-3}$ behaviour at higher energies as a result of pair cascade \cite{sve87}. The latter part of the pair energy distribution is responsible for the flat part of 
photon spectrum (in $E^2 \d N/\d E$ units). The pair cascade does not degrade the pair energies  infinitely. 
When the energy of  the photon in the jet frame drops below $1/\Thetamax\Gammaj\sim10^{4}$, it can easily escape and the cascade stops. The reduction of the mean pair energy results in an increased role of ERC, because lower-energy pairs interact with external radiation in Thomson regime. 

The photon spectra predicted by  the photon breeding model \cite{SP06,SP08,SP08b} demonstrate two-component structure (synchrotron plus Compton). The peaks shift to lower energies as luminosity increases (so called blazar sequence, \cite{fmc98}).
At low luminosities  the mean Lorentz factor of the electron-positron distribution is large (see Fig. \ref{fig:espec}), and SSC dominates over ERC because of the Klein-Nishina effect. For higher luminosities, ERC dominates the cooling.  This behaviour is in general agreement with the observational trends. However, there are some discrepancies between the shapes of the theoretical and the observed spectra. In our model at low luminosities, the  synchrotron cutoff in the 100~MeV--1~GeV range is produced by very energetic pairs which are created in the jet as a result of interaction with very soft photons. It is possible that the BLR spectrum is rather peaked, then the  injection spectrum of pairs is narrower and softer,  reducing thus the energy of the synchrotron cutoff. 

At high luminosities, our model predicts a cooling spectrum peaking around 100 MeV, while the blazars associated with quasars 
have two,  clearly separated components.  However, there is no consensus whether the two components are co-spatial.  The radio-to-optical emission in the high-luminosity sources can be produced is a separate region,  far  from the gamma-ray emitting zone.
It is also possible that the pairs produced in the jet can be heated at larger distances producing the low-energy synchrotron component.

\section{Conclusions} 

Photon breeding mechanism  can operate at various sites characterised by sufficiently  high photon density such as in the vicinity of the accretion disc,  the broad-emission line region, at parsec scale in the infrared radiation field of hot dust, at kpc scale in the stellar radiation field and even at 100 kpc scale using cosmic-microwave radiation as a target.  As a result of operation of photon breeding a decelerating, structured jets is produced.  The angular distribution of radiation is much broader than that predicted by a blob model moving with a  constant Lorentz factor.  The broad emission pattern reconciles the discrepancy between the high Doppler factors determined by the fits to
the spectra of TeV blazars and the low apparent velocities observed at VLBI scales  as well as low jet Lorentz factors 
required by the statistics and luminosity ratio of FR I  radio galaxies and BL Lac  objects. The inverse Compton emission from the pairs produced outside the jet predicts a high level of gamma-ray emission from  the misaligned jets in radio galaxies. 

Photon breeding effectively   accelerates  predominantly high-energy  leptons  in agreement with the constraints on the electron injection function determined by the spectral fits to TeV blazars.  The  mechanism also reproduces basic spectral features observed in blazars including  the blazar sequence.

\section*{Acknowledgments}
This work is supported by RFBR grant 07-02-00629-a, 
Ehrnrooth and  V\"ais\"al\"a Foundations, and  
the Academy of Finland grants 110792 and 112982.

\balance


\begin{thebibliography}{99}
%internal shocks
\bibitem{spa01} M. Spada  {\it et al.},  \Journal{{\em MNRAS}} {325} {1559}{2001}.  
%AGN jet magnetic reconnection
\bibitem{lut03}	M. Lyutikov, \Journal{{\em New Astron. Rev.}} {47} {513}{2003}. 
% =Rescattered radiation in the broad line region
\bibitem{s94} %  M. C. Begelman, M. J. Rees, 
M. Sikora  {\it et al.},  \Journal{{\em ApJ}} {421} {153}{1994}. 
%On the Efficiency of Internal Shocks in Gamma-Ray Bursts
\bibitem{amb00} A.~M. Beloborodov,  \Journal{{\em ApJ}} {539} {L25}{2000}.  
% TeV flares  from PKS 2155--304
\bibitem{aha07} F. Aharonian  {\it et al.},  \Journal{{\em ApJ}} {664}{L71}{2007}.  
% variable TeV from Mrk 501
\bibitem{alb07} J. Albert   {\it et al.}, \Journal{{\em ApJ}} {669} {862}{2007}.  
\bibitem{SP06} B. E. Stern and J. Poutanen,   \Journal{{\em MNRAS} }{372}{1217}{2006}.
\bibitem{SP08} B. E. Stern  and J. Poutanen,   \Journal{{\em MNRAS} }{383}{1695}{2008}.
\bibitem{SP08b} B. E. Stern  and J. Poutanen,   {\em Int. J. Mod. Phys. }  D, in press.
\bibitem{der03} E. V. Derishev  {\it et al.}, \Journal{{\em Phys. Rev.} D. } {68} {043003}{2003}.
%F. A. Aharonian, V. V. Kocharovsky, Vl. V. Kocharovsky, 
\bibitem{st03} B. E. Stern, \Journal{{\em MNRAS} }{345}{590}{2003}.
\bibitem{Lev96} A. Levinson, \Journal{{\em ApJ}}{467}{546}{1996}.
% decelerating jet in TeV blazars 
\bibitem{gk03} M. Georganopoulos, D. Kazanas, \Journal{{\em ApJ}} {594} {L27}{2003}. 
% structured jet in TeV BL Lac objects
\bibitem{gtc05}  G. Ghisellini, F. Tavecchio, M. Chiaberge,  \Journal{{\em A\&A}} {432} {401}{2005}. 
% beaming of ERC radiation
\bibitem{d95} C. D. Dermer,   \Journal{{\em ApJ}} {446} {L63}{1995}.    
% gamma-rays from radio galaxies and angular distribution from ERC % 
\bibitem{sso03}  {\L.} Stawarz  {\it et al.},  2003, \Journal{{\em ApJ}} {597} {186}{2003}.
% slow VLBI  jet in TeV blazars  
\bibitem{mar99} A. P. Marscher,  \Journal{{\em Astropar. Phys.}} {11} {19}{1999}.  
% The Parsec-Scale Structure and Jet Motions of the TeV Blazars 1ES 1959+650, PKS 2155-304, and 1ES 2344+514
\bibitem{pe04} B.~G. Piner and P.~G. Edwards, \Journal{{\em ApJ}} {600} {115}{2004}.   
%VLBA Polarization Observations of Markarian 421 after a Gamma-Ray High State 
\bibitem{pe05} B.~G. Piner and P.~G. Edwards, \Journal{{\em ApJ}} {622} {168}{2005}.    
% unification of radio quasars 
\bibitem{up95} C. M. Urry and P. Padovani,  \Journal{{\em PASP}} {107} { 803}{1995}. 
% Does the unification of BL Lac and FR I radio galaxies require jet velocity structures?
\bibitem{ccc00} M. Chiaberge   {\it et al.},  \Journal{{\em A\&A}} {358}{104}{2000}.   
% Unifying B2 with BL Lacs  
\bibitem{har03} M. J. Hardcastle  {\it et al.}, \Journal{{\em MNRAS}} {338} {176}{2003}.  
% delta-crisis  
\bibitem{tav05} F. Tavecchio,  2005, in Novello M.,  Perez Bergliaffa S., Ruffini R., eds,
The Tenth Marcel Grossmann Meeting. 
World Scientific Publishing, Singapore, p.\ 512
% off-axis emission in converter mechanism % F.~A. Aharonian, V.~V. Kocha\-rovsky, 
\bibitem{der07} E.~V. Derishev   {\it et al.},   \Journal{{\em ApJ}} {655} {980}{2007}.   
%, F. A. Aharonian, V. V. Kocharovsky, 
%The radio to X-ray spectrum of the M87 jet and nucleus % C. P. Stern, D. E. Harris,
 \bibitem{bsh91}  A. Biretta  {\it et al.},  \Journal{{\em AJ}} {101} {1632}{1991}.   
%The Supermassive Black Hole of M87 and the Kinematics of Its Associated Gaseous Disk
\bibitem{mac97} F. Macchetto  {\it et al.},  \Journal{{\em ApJ}} {489} {57}{1997}.     
% TeV from M87
\bibitem{aha06} F. Aharonian  {\it et al.},  \Journal{{\em Sci}} {314}{1424}{2006}.  
% limb brightening in Mrk 501 at VLBI pc scale 
\bibitem{gir04} M. Giroletti  {\it et al.},  \Journal{{\em ApJ}} {600} {127}{2004}.  
% fits to the blazars spectra %  A. Celotti, L. Costamante, 
\bibitem{gcc02} G. Ghisellini {\it et al.}, \Journal{{\em A\&A}} {386} {833}{2002}. 
% fit to Mrk 501 %P. S. Coppi, F. Aharonian,
\bibitem{kca02}	 H. Krawczynski   {\it et al.},  \Journal{{\em MNRAS} }{336}{721}{2002}.
% fit to Mrk 421 & 501 
\bibitem{kon03} A. Konopelko  {\it et al.}, \Journal{{\em ApJ}} {597} {851}{2003}.  
 % fits to TeV blazars   % , G. Dubus, B. Kh\'elifi
\bibitem{gdk07} B. Giebels  {\it et al.}, \Journal{{\em A\&A}} {462} {29}{2007}. 
 % pair cascade
\bibitem{sve87} R. Svensson, \Journal{{\em MNRAS} }{227}{403}{1987}. 
% blazar sequence - data 
\bibitem{fmc98} G. Fossati  {\it et al.}, \Journal{{\em MNRAS}} {299} {433}{1998}.

\end{thebibliography}
\end{document}